# Observation of a Logarithmic Temperature Dependence of Thermoelectric Power on Multiwall Carbon Nanotubes

N. Kang, L. Lu*, W. J. Kong, J. S. Hu, W. Yi†, Y. P. Wang, D. L. Zhang, Z. W. Pan‡, and S. S. Xie

*Key Laboratory of Extreme Conditions Physics, Institute of Physics & Center for Condensed Matter Physics*
*Chinese Academy of Sciences, Beijing 100080, P. R. China*

We have investigated the thermoelectric power (TEP) of millimeter-long aligned multiwall carbon nanotubes down to a temperature of $T=1.5$ K, and observed for the first time an accurate $T\ln T$ dependence at low temperatures. This behavior is possibly originated from the repulsive interaction between the electrons in a disordered local environment.

The discovery of carbon nanotube [1] provides us an intriguing system to study correlated electrons at low dimensions. In a single-wall nanotube (SWNT) it is clear that the electrons are highly correlated to form a one-dimensional (1D) Luttinger Liquid (LL) [2] and that their transport is via ballistic processes [3–6]. In a multiwall nanotube (MWNT), however, the case is not so clear. Inter-wall coupling and Coulomb screening tend to break down the unconventional LL picture and draw the electrons back to the conventional Fermi liquid (FL). The scattering processes of the electrons by defects (such as the intrinsic stacking faults between adjacent walls [7]) should also influence the transport properties [8], leading to a diffusive rather than a ballistic electron motion. Experimentally, the magneto-transport measurements [9,10] did reveal that the motion of electrons in a MWNT is diffusive, showing the tendency of weak localization similar to a 2D disordered FL. Nevertheless, the electron system in a MWNT seems still unconventional, as that a power-law-like tunneling density of states [11–14] similar to that of the LL is preserved. To further understand such an interesting electron system, experiments other than tunneling and magnetotransport measurements are needed.

TEP measurement can provide us very useful information of the electrons. Previously, a positive and roughly linear temperature dependent TEP has been observed on mats of SWNTs by J. Hone et al. [15] and on mats of MWNTs by M. Tian et al. [16]. A pronounced TEP peak at 100 K has been observed on some transition-metal doped SWNTs mats by L. Grigorian et al. [17]. And a much linear temperature dependence has been observed on a single MWNT sample by P. Kim et al. [18]. In each of these experiments, no anomalies could be recognized as $T \to 0$. In this letter, we report our TEP investigation on bundles of MWNTs. With an improved signal-to-noise ratio, we found that the TEP of our MWNTs is clearly suppressed below 20 K — it changes from a linear to a $T\ln T$ dependence. To our knowledge, this is the first observation of such a logarithmic behavior of TEP. We will discuss the possible origins of this phenomenon.

The samples used in this experiment are bundles of MWNTs grown by chemical vapor deposition [19]. The MWNTs in the bundle are about 30 nm in diameter and reach millimeters in length. They are aligned in parallel and separated from each other at a distance of $\sim 100$ nm except for occasional points of contact which should not affect the TEP measurement [20]. We used an ac method to measure the TEP. In this method the temperature fluctuation across the sample was created by feeding into a mini-heater a 2 kHz ac current whose amplitude was slowly modulating at a few Hertz. A lock-in amplifier was used to detect the TEP signal of the sample at the modulation frequency. In this way, the 2 kHz voltage component and most of the noise were filtered out, and a resolution of 20 nV/K was reached. With this technique, previously we were able to measure the TEP of decagonal quasicrystals very accurately [21]. To further guarantee the absolute accuracy of our measurements at low temperatures, we used a stripe of $Bi_2Sr_2CaCu_2O_{8+x}$ high-$T_c$ superconductor as the counter arm below $\sim 75$ K (illustrated in Fig. 1).

The upper panel of Fig. 1 shows the temperature dependence of MWNTs' TEP from 1.5 to 300 K [22]. Six MWNT samples have been investigated and the results are the same: the TEP is positive, increasing with temperature and reaching $\sim +30$ $\mu$V/K at room temperature. This behavior is qualitatively similar to the earlier observations on SWNT and MWNT samples [15,16,18].

Compared to the earlier experiments, however, the high signal-to-noise ratio we reached allowed us to distinguish a downward deviation from the linear temperature dependence below 20 K. The deviation can be clearly seen in Fig. 2, by plotting $S/T$ against $\log_{10} T$, where $S$ denotes the TEP. The data line up in accordance with the following logarithmic law:

$$\frac{S}{T} = \alpha - \beta \ln\left(\frac{T_0}{T}\right) \quad (1)$$

where $T_0 \approx 20$ K, $\alpha \approx 0.15 \mu$V/K$^2$, and $\beta \approx 0.10 \mu$V/K$^2$.

Using the same setup and method, we also checked the TEP of glassy carbon and graphite crystal samples for comparison. The glassy carbon sample was a bar-like piece taken from a Speer carbon resistor, and the crystal sample was a stripe of HOPG graphite. The results, as shown in the lower panel of Fig. 1, are consistent with the other reports [16,23]. The inset of this panel shows that the TEP of both glassy carbon and graphite crystal varies linearly with T as $T \to 0$, confirming that the low-temperature suppression in TEP is a particular property of the MWNT samples.

The MWNT bundles used in this experiment contain several hundred individual MWNTs. Each MWNT consists of several ten co-axial shells. Although the outmost shell is believed to be the most important contributors to the transport measurements, there is recent evidence that the inner ones also contribute substantially [24]. About one-third of the undoped tubes are metallic at low temperatures [25] and the other two-thirds are semiconducting, with a gap of $\Delta E = \hbar v_F / d \sim 10$ meV for tubes of diameter $d = 30$ nm like ours (where $\hbar$ is Planck's constant, and $v_F$, the Fermi velocity). 10 mV is also the energy scale of the subbands separation near the Fermi level for both our metallic and semiconducting tubes. It is known that the TEP of carbon nanotubes is sensitive to gas adsorption [26,27]. A MWNT that has ever been exposed to air is doped with holes so that it has several residual conduction channels. At temperatures above 10 meV, namely $\sim 100$ K, the subbands above the Fermi level will contribute to our measurement due to thermal activation. Varying temperatures will therefore change the number of conduction channels and thus influence the temperature dependence of the TEP. Nevertheless, the number of channels should be well fixed at $T \ll 100$ K. Therefore, the observed $T\ln T$ dependence in TEP should be safely regarded as an intrinsic property of the multiple-channel electron system.

Previously it has been found that the magnitude and the temperature-dependent shape of SWNTs' TEP depend sensitively on the type of transition-metal catalysts used in the synthesis, showing a pronounced peak due to the Kondo mechanism [17]. As our MWNTs were grown with iron catalysts, it is interesting to see if the TEP is influenced by a similar mechanism. To address this issue, we notice that the Kondo mechanism causes an enhancement to the TEP, whereas the TEP of our MWNTs undergoes a suppression from the linearity below 20 K. This suppression is in a dominant form of $T\ln T$ which can not explained, to our knowledge, with the existing theories of Kondo effect [28]. In addition, according to the Gorter-Nordheim rule [29], the TEP signal due to the Kondo mechanism should be largely suppressed by the numerous non-magnetic scattering processes existed in our MWNTs. We also note that no iron atoms can be identified in the body of our MWNTs in the resolution limits by energy dispersion X-ray spectroscopy and transmission electron microscopy studies. At this moment, therefore, we could not identify Kondo effect as the origin of the $T\ln T$ behavior.

In the following we will show that the logarithmic suppression in TEP is possibly a combined effects of electron-electron (e-e) interaction and electron-disorder scattering in the MWNT. As mentioned in the introduction, the MWNT contains both intrinsic and extrinsic defects. The existence of defects in our MWNTs is confirmed by our previous Raman scattering and thermal conductivity measurements [30]. With these defects, the electrons will likely fall into weak localization (WL) at low temperatures [31,32]. They may enter into a 1D WL state if their dephasing length is longer than the circumference of the tube, or enter into a 2D WL state if otherwise. Some of the earlier electron transport measurements confirmed the 2D WL in MWNT [9,10].

There has been great interest in weakly disordered electron systems over the past few decades [31,32]. It has been predicted that the e-e Coulomb interaction will be enhanced in the presence of defect scattering, resulting in corrections on the electron's density of states and on the transport coefficients such as conductivity, magneto-conductivity and TEP. In particular, the temperature dependent conductivity and magneto-conductivity have been calculated and fitted to numerous experimental data with great success. For TEP, the influence of e-e interaction on 2D WL electrons was first considered by C. S. Ting et al. [33] two decades ago. By introducing the e-e interaction as a perturbation, a logarithmic correction on the temperature dependence of TEP was predicted:

$$S = S_0 \left[1 - \frac{1}{2\pi E_F \tau}\left(1 - \frac{3}{2}F\right)\ln\left(\frac{1}{T\tau}\right)\right] \qquad (2)$$

$$S_0 = \frac{\pi^2 k_B^2 T}{3e}\left[\frac{\partial \ln\sigma}{\partial E}\right]_{E_F} \qquad (3)$$

where Eq. (3) is Mott's formula of TEP for diffusive non-interacting electrons, $E_F$ is the Fermi energy, $\tau$, the electron momentum relaxation rate due to impurity scattering, $F$, the measure of Coulomb screening, and $\sigma$, the electrical conductivity. A logarithmic correction on TEP was also predicted by J. W. P. Hsu et al. [34] and by K. D. Belashchenko et al. [35]. However, no experimental evidence has previously been reported.

Eq. (2) is equivalent to Eq. (1) which describes our experimental data below 20 K. With a typical Fermi energy $E_F \sim 0.1$ eV and a relaxation rate $\tau = 1 \sim 2 \times 10^{-14}$ S as obtained from our magnetoresistance measurement [36], the TEP data below 20 K yield a reasonable screen constant of $F \approx 0.3$. Therefore, the above theories of 2D WL might approach the underlying physics of the $T\ln T$ behavior of TEP, at least at relatively high temperatures.

A dominant logarithmic behavior at low temperatures is, however, more than one would expected based on the above perturbation theories. Considering that some of the LL features such as a power-law-like zero bias anomaly in tunneling conductance is preserved in MWNT, it would be interesting to ask if one could approach the $T\ln T$ law using an unconventional picture of electrons. Indeed, a dominant logarithmic law of TEP would imply that the electron system is near to some critical point where no particular energy scale exists.

Along this line of seeking for unconventional and possibly better understandings to the experimental data, let us mention that Kane and Fisher [37] has already predicted a linear temperature dependence of TEP for defect-free 1D LL. Therefore, LL alone cannot account for the logarithmic TEP suppression.

Recently, the peculiar disordered electron system in MWNT has been studied by taking the e-e interaction non-perturbatively. In their theory explaining the logarithmic and power-law-like zero bias anomaly in MWNT, Egger and Gogolin [38] have shown the importance of a nonconventional Coulomb blockade mechanism on the diffusive motion of the electrons. This mechanism should also impede the electrons from accumulating along the temperature gradient, resulting in suppression in TEP. In another non-perturbative calculation on the tunneling density of states of an N-channel disordered wire, Mishchenko and co-workers [39] predicted the formation of a Coulomb gap at low energies. The TEP suppression we observed is also consistent qualitatively with this prediction, because in a disordered metal such a gap will again prevent the electrons from redistributing along the temperature gradient. The temperature-dependent conductance $G$ of our MWNTs seems to follow the form of $-\ln G \propto T^{-1/2}$ at low temperatures (Fig. 3), which supports the opening up of a Coulomb gap [39,40]. It will be interesting to see if the logarithmic law of TEP can be explicitly deduced from either of the above non-perturbation theories.

In conclusion, we found that the TEP of the MWNT has a logarithmic temperature dependence at low temperatures. This behavior is likely resulted from the interplay between electron-electron Coulomb repulsion and electron-disorder scattering. Further theoretical and experimental investigations are needed to fully clarify the intriguing multiple-channel electron system in MWNT.

## ACKNOWLEDGMENTS

We thank T. Xiang, R. Egger, G. M. Zhang, Y. P. Wang and Q. Niu for helpful discussions and sharing of results prior to publication. This work is supported by the National Science Foundation of China.

* correspondence author, email: *lilu@aphy.iphy.ac.cn*.
† present address: Department of Applied Physics, Harvard University, Cambridge, Massachusetts 02138, U. S. A.
‡ present address: School of Materials Science and Engineering, Georgia Institute of Technology, Atlanta, Georgia 30332, U. S. A.

FIG. 1. Upper Panel: Thermoelectric power of multi-wall carbon nanotube samples as a function of temperature. The upper-left inset is the schematic for this measurement. The data above 75 K were measured using a gold wire of 0.5 milli-inch in diameter as the counter arm, whereas below 75 K the gold wire was replaced by a thin stripe of $Bi_2Sr_2CaCu_2O_{8+x}$ high-$T_c$ superconductor. The suppression of TEP from linearity at low temperatures is clearly shown in the lower-right inset. Lower Panel: The thermoelectric power of HOPG graphite single crystal and glassy carbon samples. (Inset) No suppression was observed for either sample as $T \to 0$.

FIG. 2. Low-temperature thermoelectric power of multi-wall carbon nanotube samples plotted as $S/T$ on a logarithmic temperature scale, showing a well-defined logarithmic suppression below $\sim 20$ K. The plotted data were measured on two different samples.

FIG. 3. The conductance $G$ of the multi-wall carbon nanotube samples as a function of temperature. It crosses from a logarithmic-like law to a power-like law of the form $-\ln G \sim T^{-1/2}$ at low temperatures. The deviation of $G$ from the logarithmic behavior above $\sim 100$ K is due to the contributions of the high subbands.

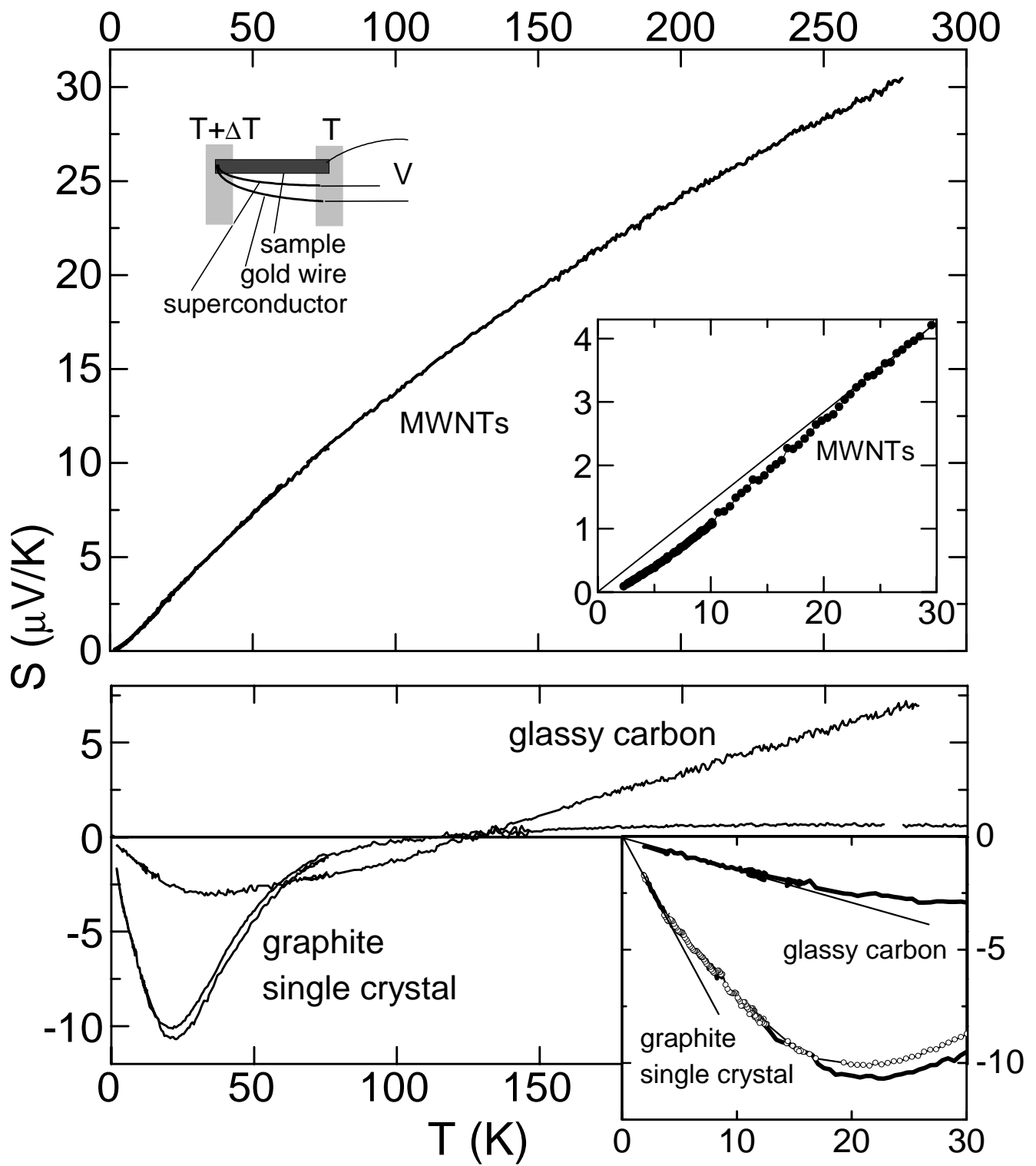

Kang, et al., Fig. 1

Upper Panel: Thermoelectric power of multi-wall carbon nanotube samples as a function of temperature. The upper-left inset is the schematic for this measurement. The data above 75 K were measured using a gold wire of 0.5 milli-inch in diameter as the counter arm, whereas below 75 K the gold wire was replaced by a thin stripe of ${\rm Bi_2Sr_2CaCu_2O_{8+x}}$ high-T$_c$ superconductor. The suppression of TEP from linearity at low temperatures is clearly shown in the lower-right inset. Lower Panel: The thermoelectric power of HOPG graphite single crystal and glassy carbon samples. (Inset) No suppression was observed for either sample as $T\to 0$.

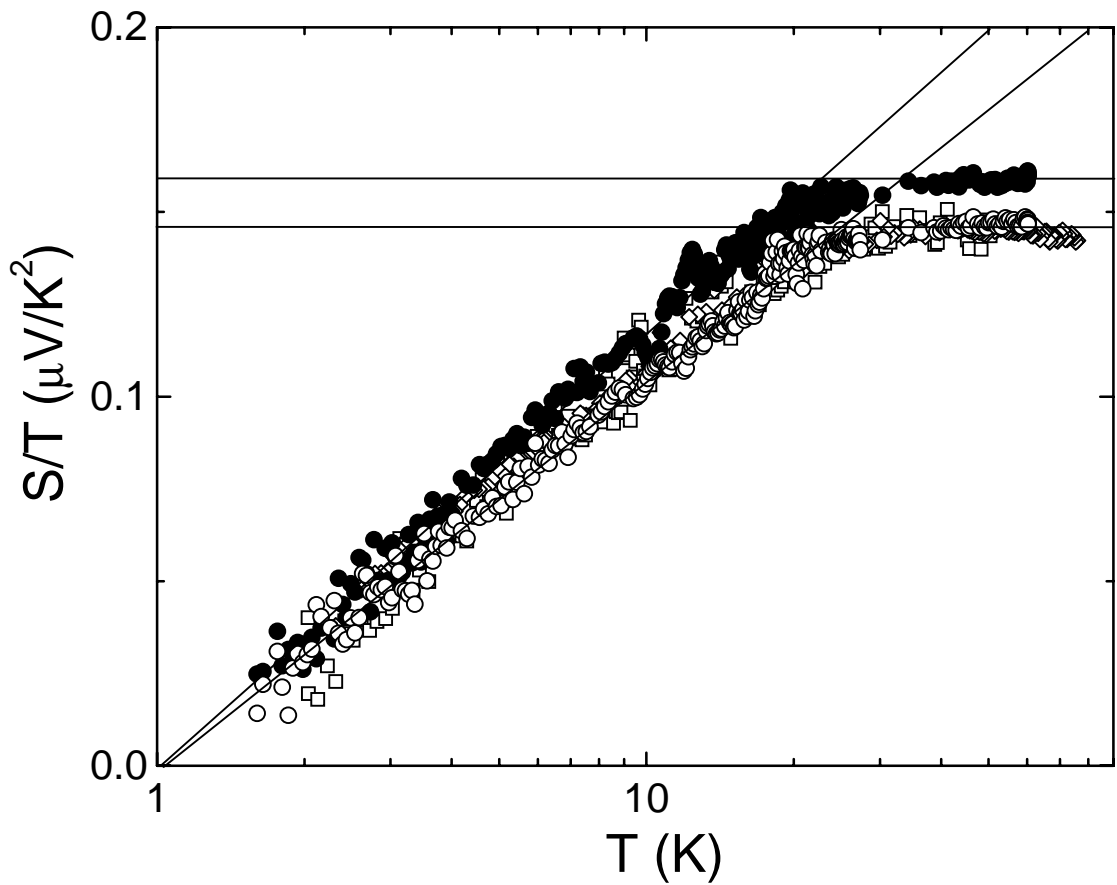

N. Kang, et al., Fig. 2

Low-temperature thermoelectric power of multi-wall carbon nanotube samples plotted as $S/T$ on a logarithmic temperature scale, showing a well-defined logarithmic suppression below $\sim$ 20 K. The plotted data were measured on two samples in different temperature runs.

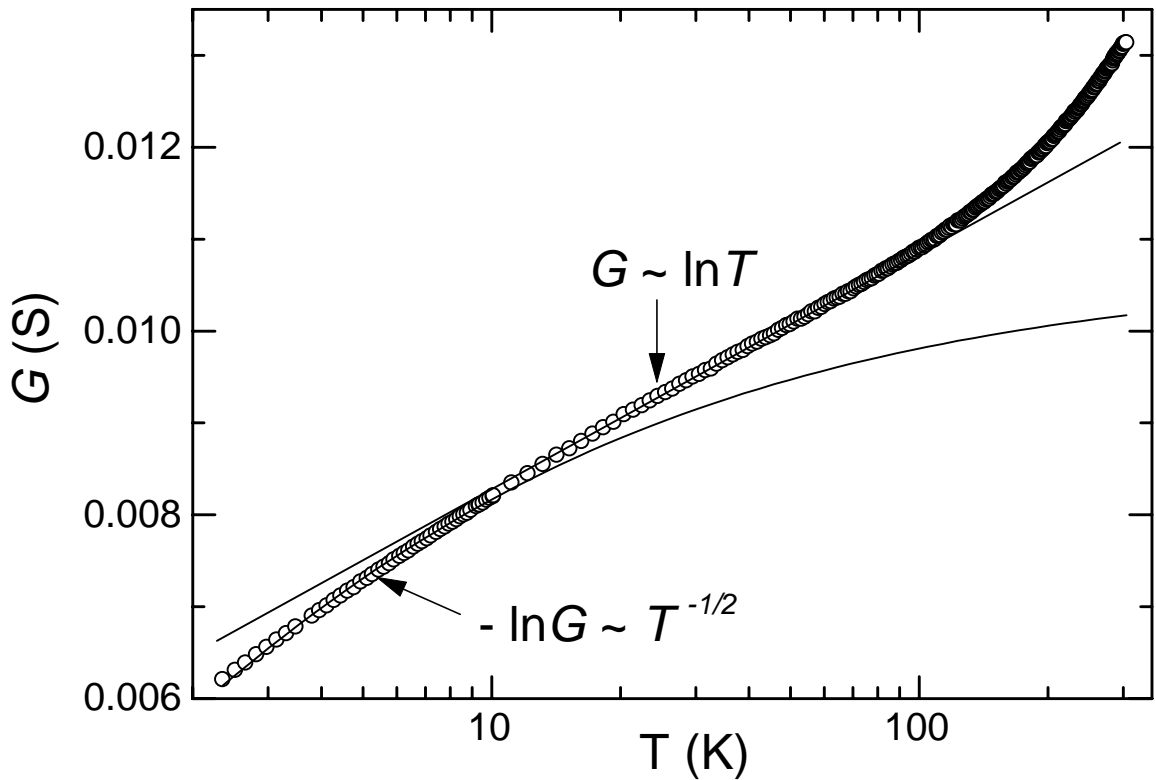

Kang et al. Fig. 3

The conductance $G$ of the multi-wall carbon nanotube samples as a function of temperature. It crosses from a logarithmic-like law to a power-like law of the form $-{\rm ln} G \sim T^{-1/2}$ at low temperatures. The deviation of $G$ from the logarithmic behavior above $\sim$ 100 K is due to the contributions of the high subbands.